\documentclass[aps,letterpaper,nofootinbib,preprint,showpacs,preprintnumbers,amsmath,amssymb]{revtex4}%
\usepackage{latexsym}%
\usepackage{epsf}%
\usepackage[hypertex]{hyperref}%

\def\oP{{\overline P}}

\def\tR{{\widetilde R}}
\def\tF{{\widetilde F}}
\def\tG{{\widetilde G}}
\def\tZ{{\widetilde Z}}
\def\tbeta{{\widetilde \beta}}

\def\pA{{\mathpzc A}}
\def\cF{{\mathcal F}}
\def\cH{{\mathcal H}}

\def\cN{{\mathcal N}}

\def\cQ{{\mathcal Q}}
\def\cW{{\mathcal W}}

\DeclareMathAlphabet{\mathpzc}{OT1}{pzc}{m}{it}

\newcommand{\beq}{\begin{equation}}
\newcommand{\beqn}{\begin{equation}\nonumber}
\newcommand{\eeq}{\end{equation}}
\newcommand{\bea}{\begin{eqnarray}}
\newcommand{\bean}{\begin{eqnarray}\nonumber}
\newcommand{\eea}{\end{eqnarray}}

\begin{document}

\begin{center}
{\bf{\Large Spectrum and Statistical Entropy of AdS Black Holes}}
\bigskip
\bigskip

{{Cenalo Vaz$^{a,b,}$\footnote{e-mail address: Cenalo.Vaz@UC.Edu},
L.C.R. Wijewardhana$^{b,}$\footnote{e-mail address: Rohana.Wijewardhana@UC.Edu}}}
\bigskip

{\it$^a$RWC and $^b$Department of Physics,}\\
{\it University of Cincinnati,}\\
{\it Cincinnati, Ohio 45221-0011, USA}
\medskip

\end{center}
\bigskip
\bigskip
\medskip

\centerline{ABSTRACT}
\bigskip\bigskip
Popular approaches to quantum gravity describe black hole microstates differently and
apply different statistics to count them. Since the relationship between the approaches 
is not clear, this obscures the role of statistics in calculating the black hole entropy. 
We address this issue by discussing the entropy of eternal AdS black holes in dimension 
four and above within the context of a midisuperspace model. We determine the black 
hole eigenstates and find that they describe the quantization in half integer units of 
a certain function of the Arnowitt-Deser-Misner (ADM) mass and the cosmological constant. 
In the limit of a vanishing cosmological constant (the Schwarzschild limit) the quantized 
function becomes the horizon area and in the limit of a large cosmological constant it 
approaches the ADM mass of the black holes. We show that in the Schwarzschild limit the 
area quatization leads to the Bekenstein-Hawking entropy if Boltzmann statistics are 
employed. In the limit of a large cosmological constant the Bekenstein-Hawking entropy 
can be recovered only via Bose statistics. The two limits are separated by a first order 
phase transition, which seems to suggest a shift from ``particle-like'' degrees of freedom 
at large cosmological constant to geometric degrees of freedom as the cosmological 
constant approaches zero.

\bigskip

%%%%%%%%%%%%%%%%%%%%%%%%%%%%%%%%%%%%%%%%%%%%%%%%%%%%%%%%%
	\vfill\eject
\section{Introduction\label{intro}}

According to the results of Bekenstein, Hawking and others \cite{bek72,bek72a,bek73,bch73,haw75}, 
a matter cloud of one solar mass would gain entropy by a factor of roughly $10^{20}$ in collapsing 
to form a black hole. The degrees of freedom resposible for this dramatic increase in entropy are 
attributed to quantum gravity. About a decade ago it was believed that a successful microcanonical 
derivation of the  Bekenstein-Hawking entropy of a black hole would strongly support the particular 
theory of quantum gravity from which the black hole microstates were derived, but subsequent 
research has shown that various seemingly different approaches to quantum gravity successfully 
reproduce the black hole entropy from a microcanonical or a canonical ensemble. However there 
are significant variations in the details of the calculations.

Three of the most popular candidates for black hole microstates are duals of weak field 
string and D-brane states in string theory \cite{stva96,hrpl97,dab05,math05,math08}, states of a 
boundary conformal field theory (CFT) \cite{hkl99,emp99,ahaoo00,hmst01,mupa02,chgs07} and punctures 
of a spin network on the horizon in loop quantum gravity \cite{rov96,abck97,dle04,mei04}. 
The microstates being counted differ between approaches and there are also differences in the 
statistics employed to count them. In the string and AdS/CFT approaches, the microstates are 
counted using Bose statistics, but in loop quantum gravity they are counted using Boltzmann 
statistics. In fact, if the loop states are assumed indistinguishable one obtains an entropy 
that is proportional to the square root of the horizon area \cite{Kolland}. It is quite likely 
that the ability of different approaches to reproduce the Bekenstein-Hawking entropy is suggestive 
of a ``universality'' of black hole entropy as proposed in \cite{carlip08}, but it is 
nevertheless of interest to understand the application of different statistics in the 
counting of the microstates better. Unfortunatley, because the relationships between approaches 
are poorly understood, it has been difficult to address this issue until now.

Here we discuss the role of statistics in the context of a single model of the black hole, applying the 
results of a midisuperspace quantization program describing the spherical collapse of inhomogeneous 
dust in any number of spatial dimensions both with and without a negative cosmological constant
\cite{va3,va4}. While the model is not derived from any deeper theory of quantum gravity such as 
string theory or loop quantum gravity, it is reasonable to expect that it can adequately address at 
least the semi-classical features of black holes. We ask what statistics are required to recover the
leading behavior of the black hole entropy. 

The black hole eigenstates are described by the quantization in half integer units of a certain function,
$\pA(M,l)$, of its mass and the cosmological constant ($\Lambda = -l^{-2}$). If we set $x_h = R_h/l$, 
where $R_h$ is the area radius of the horizon, we find that when $x_h^2 \ll 2n(n+1)$ the quantity 
$\pA(M,l)$ approaches the horizon area. The Schwarzschild black hole belongs in this region and we 
shall refer to this as the Schwarzschild limit. Thus it is the area of the Schwarzschild horizon that 
admits a linear spectrum, in keeping with Bekenstein's original proposal \cite{bek72}. On the other 
hand when $x_h^2 \gg 2n(n+1)$, the quantity $\pA(M,l)$ approaches the ADM mass and it is the black hole 
{\it mass} that admits the same linear spectrum. 
We show that, to recover an area dependence coinciding with the Bekenstein-Hawking entropy (to leading 
order, modulo fluctuations), one must use Boltzmann statistics in the first limit and Bose statistics 
in the second limit. Further, the Bose partition function is dominated by a contribution from the 
Schwarzschild limit, which enters via a well known duality linking its high temperature behavior to its 
low temperature dynamics \cite{hr18}.

In section II we briefly recall some of the basic results from the canonical formulation of $d=2+n$ 
dimensional spherically symmetric gravity as in \cite{va4}. For details of the classical solutions 
and their behavior we refer the reader to \cite{va3}. In section III we obtain stationary bound 
states describing (eternal) black holes and then, in Section IV we use the black hole spectrum to 
determine the statistical entropy in the two limiting cases described in the previous paragraph.
We conclude in Section V.

\section{Canonical Formulation}

We have already described the 2+1 dimensional black hole in detail \cite{va2}, so here we will 
concentrate on the case $n\geq 2$. The models are represented by solutions of Einstein's equations 
sourced by the stress tensor $T_{\mu\nu} = \varepsilon U_\mu U_\nu$, where $\varepsilon(t,\rho)$ 
is the dust proper energy. In any number of dimensions and in comoving coordinates, the solutions 
are characterized by two arbitrary functions of the radial coordinate $\rho$, {\it viz.,} the 
``mass function'', $F(\rho)$, and the ``energy function'', $E(\rho)$. In comoving coordinates, 
$\rho$ serves as a spatial label for collapsing dust shells. The mass function $F(\rho)$ represents 
the mass-energy contained within a dust shell labeled by $\rho$ and the energy function is related 
to the initial velocity distribution of the shells.  The classical solutions are given by
\beq
ds^2 = d\tau^2 - \frac{\tR^2}{1+2E} d\rho^2 + R^2 d\Omega_n^2
\label{metric}
\eeq
where $\Omega_n$ is the solid angle for the $n-$sphere, $R(\tau,\rho)$ is the radius of the 
$n-$sphere and we have used the notation $\tR(\tau,\rho)=\partial_\rho R(\tau,\rho)$. Einstein's 
equations then give
\bea
&&\varepsilon(\tau,\rho) = \frac{(n-1)}{8\pi G_d} \frac{\tF}{R^n\tR},\cr\cr
&&{R^*}^{2}=-\frac{2\Lambda}{n(n+1)}R^{2}+\frac{F(\rho)}{R^{n-1}} + E(\rho)
\label{eqmot}
\eea
where $G_d$ is the $d-$dimensional gravitational constant, $\widetilde{F} = \partial_\rho 
F(\rho)$ and $R^*(\tau,\rho) = \partial_\tau R(\tau,\rho)$. Shells labeled by $\rho$ become 
singular as the area radius approaches zero. These classical solutions have been analyzed 
in \cite{va3}. 

The general spherically symmetric ADM metric 
\beq
ds^2 = N^2 dt^2 - L^2 (dr+N^r dt)^2 +R^2d\Omega_n^2
\eeq
can now be embedded in the spacetime described by \eqref{metric}. This procedure was shown in 
\cite{va4} to give a canonical description of the collapse in terms of a phase space consisting 
of the dust proper time, $\tau(r)$, the area radius, $R(r)$, and the mass density function, 
$\Gamma(r)$, which is defined in terms of the mass function via
\beq
F(r) = \frac{\tG_d}n M_0 + \frac{\tG_d}n\int_0^r \Gamma(r') dr',
\label{massfn}
\eeq
where $\tG_d = 16\pi G_d/\Omega_n$ and $M_0$ represents the boundary contribution to the 
hypersurface action at the center. In dimension four and higher the boundary contribution from 
the origin is generally set to zero because a non-vanishing mass function at the origin would 
represent a singular initial configuration corresponding to a point mass at the 
center\footnote{The situation is different in 2+1-dimensions. A non-vanishing contribution from 
the origin is essential to allow for an initial velocity profile that vanishes there. This does 
not lead to singular  initial data and the boundary contribution does not have the interpretation 
of a point mass situated at the center.}, so we restrict our attention to $M_0=0$. As shown in 
\cite{va4}, the boundary terms can be absorbed into a single hypersurface action and the 
effective constraints of the gravity dust system reduce to
\bea
&&\cH_r = \tau'P_\tau + R' \oP_R -\Gamma P_\Gamma' \approx 0\cr\cr
&&\cH^{g}=P_\tau^2 + \cF\oP_R^2 -\frac{\Gamma^2}{\cF}\approx 0,
\label{const1}
\eea
where the prime refers to a derivative with respect to the ADM label coordinate $r$, and 
\beq
\cF = 1-\frac F{R^{n-1}}+\frac{2\Lambda R^2}{n(n+1)}
\label{cF}
\eeq
vanishes on the apparent horizon, passing from negative inside to positive outside. The first 
of \eqref{const1} represents the diffeomorphism (momentum) constraint and the second is the 
Hamiltonian constraint. This simplified form of the constraints was obtained after several 
canonical transformations on the original (ADM) phase space in the spirit of Kucha\v r 
\cite{kuc94}, squaring the Hamiltonian of the transformed system and imposing the momentum 
constraint. Because of the absence of any derivative terms in the Hamiltonian constraint, it 
is much easier to quantize than the original gravity-dust system and we use it as the starting 
point for our discussion. In the quantum theory, the apparent horizon is treated as a 
boundary at which both continuity and differentiability of the wave functional which solves the 
Wheeler-DeWitt equation are required.

\section{Quantum States}

Dirac's quantization procedure may now be applied to turn the classical constraints into 
operator constraints on wave-functionals. According to it, the momenta are replaced by 
functional differential operators (we set $\hbar=1$)
\beq
{\widehat P}_X = -i  \frac{\delta}{\delta X(r)},
\label{diracq}
\eeq
and one may write the quantum Hamiltonian constraint as \cite{k3}
\beq
{\widehat \cH} \Psi[\tau,R,\Gamma] = \left[\frac{\delta^2}{\delta \tau^2} + 
\cF \frac{\delta^2}{\delta R^2} + A\delta(0) \frac{\delta}{\delta R} + B\delta(0)^2+
\frac{\Gamma^2}{\cF}\right] \Psi[\tau,R,\Gamma] = 0,
\label{qham}
\eeq
where $A(R,F)$ and $B(R,F)$ are smooth functions of $R$ and $F$ which encapsulate the factor 
ordering ambiguities. The divergent quantities $\delta(0)$ and $\delta(0)^2$ are introduced 
to indicate that the factor ordering problem can be dealt with only after a suitable 
regularization procedure has been implemented. We notice that the Hamiltonian constraint 
contains no functional derivative with respect to the mass density function. In fact the mass 
density appears merely as a multiplier of the potential term in the Wheeler-DeWitt equation. 
This indicates that $\Gamma(r)$, and hence the initial energy density distribution, 
$\varepsilon(0,r)$ may be externally specified. Once specified, $\Gamma(r)$ determines the 
quantum theory of a particular classical model. 

The quantum momentum constraint, on the other hand, 
\beq
{\widehat \cH}_r \Psi[\tau,R,\Gamma] = \left[\tau'\frac{\delta}{\delta \tau} + R'\frac{\delta}
{\delta R} - \Gamma \left(\frac{\delta}{\delta \Gamma}\right)' \right] \Psi[\tau,R,\Gamma] = 0,
\label{qdiff}
\eeq
requires no immediate regularization because it involves only first order functional derivatives.
To describe a collapsing cloud with a smooth, non-vanishing matter density distribution over 
some label set, $r$, of non-zero measure the Hamiltonian constraint was regularized on a lattice. 
Accounting for geometric factors and assuming that the wave-functional is factorizable, the 
continuum limit of the wave-functional can quite generally be taken as
\beq
\Psi[\tau,R,\Gamma] = \exp\left[i\int dr \Gamma(r) \cW(\tau(r),R(r),F(r))
\right].
\label{wfnal}
\eeq
It automatically obeys the momentum constraint provided that $\cW(\tau,R,F)$ has no explicit 
dependence on the label coordinate $r$. 

We showed in \cite{k3} that, for the wave-functionals to be simultaneously factorizable on the 
lattice and to obey the momentum constraint in the continuum limit (as the lattice spacing is 
made to approach zero), they must satisfy not one but three equations,
\bea
&&\left(\frac{\partial W}{\partial\tau}\right)^{2}+\mathcal{F}\left(\frac{\partial W}{\partial\ R}
\right)^{2}-\frac{4}{\mathcal{F}}=0,\cr\cr
&&\left(\frac{\partial^{2}}{\partial\tau^{2}}+\mathcal{F}\frac{\partial^{2}}{\partial R^{2}}+A
\frac{\partial}{\partial R}\right)W=0,\cr\cr
&&B=0,
\label{eqns}
\eea
the first of which is the Hamilton--Jacobi equation that was used in earlier studies \cite{k1} 
to describe Hawking radiation in the WKB approximation. The function $B(R,F)$ in \eqref{qham} is 
forced to be identically vanishing. The remaining two equations together with hermiticity of 
the Hamiltonian constraint uniquely fix the measure, $\mu$, that defines an inner product on 
the Hilbert space, determine $A(R,F)$ in terms of this measure,
\beq
A = |\cF|\partial_{R} \ln (\mu|\cF|),
\label{meas}
\eeq
and also determine $W(\tau,R,F)$. The solutions were shown to yield Hawking radiation in 
\cite{va4}. Lattice regularization effectively turns the continuum (midi-superspace) problem 
into a countably infinite set of decoupled mini-superspace problems; the three equations 
mentioned earlier are required to ensure a sensible, diffeomorphism invariant continuum limit. 

Black holes with ADM mass parameter $M$ are special cases of the solution in \eqref{metric}, 
obtained when the mass function is constant, $F=2G_d M$, and the energy function is vanishing. 
This can be shown directly by a coordinate transformation of \eqref{metric} from the comoving 
system $(\tau,\rho)$ to static coordinates $(T,R)$, in which the metric has the standard from,
\beq
ds^2 = \cF(R)dT^2 - \cF^{-1}(R) dR^2 - R^2 d\Omega_n^2,
\label{bh2}
\eeq
where $\cF$ is given in \eqref{cF}. To show this, use the solution $R=R(\tau,\rho)$ of 
\eqref{eqmot} together with the relationship between dust proper time and Killing time
\beq
\tau = T - \int dR \frac{\sqrt{1-\cF}}\cF,
\eeq
which was obtained in \cite{va4}. We imagine therefore that the black holes 
are single shells given by the mass function
\beq
F(r) = 2 G_d M \Theta(r)
\label{massfn2}
\eeq
where $M$ is the mass at label $r=0$ and $\Theta(r)$ is the Heaviside function. The mass density function 
[see \eqref{massfn}] is therefore 
\beq
\Gamma(r) = \frac {n\Omega_n}{8\pi} M \delta(r)
\eeq
and, because of the $\delta-$distributional mass density, the wave-functional in \eqref{wfnal} turns
into the wave-{\it function}
\beq
\Psi[\tau,R,\Gamma] = e^{i\int_0^\infty dr \Gamma(r) \cW(\tau(r),R(r),F(r))}
 = e^{\frac{i n\Omega_n}{8\pi} M\cW(\tau,R,F)}
\eeq
where $\tau=\tau(0)$, $R=R(0)$ and $F=F(0)$. The Wheeler-DeWitt equation now becomes the Klein-Gordon
equation describing the shell. Taking into account the factor ordering ambiguities and absorbing 
the $M$ dependent term, which now renormalizes the potential, into the function $B(R,F)$ we have
\beq
\left[\frac{\partial^2}{\partial \tau^2}+\cF\frac{\partial^2}{\partial R^2} + A
\frac{\partial}{\partial R}+B\right]e^{\frac{i n\Omega_n}{8\pi}M\cW(\tau,R,F)} = 0,
\label{wd1}
\eeq
In contrast with the case in which the mass density is a smooth function over some set of non-zero 
measure, no regularization is necessary here. This means that no further conditions must be met
and therefore that the measure as well as the functions $A(R,F)$ and $B(R,F)$ will remain undetermined 
although the function $A(R,F)$ will continue to be related to the measure according to \eqref{meas}.
Thus two conditions are required to proceed with the quantization of the black holes as described above.
 
The first condition we impose is one on the measure appropriate to the Hilbert space of wave-functions.
In \cite{va4} we obtained the Hawking evaporation of a collapsing dust cloud surrounding a pre-existing 
black hole by taking the dust as a small perturbation to the black hole mass function in \eqref{massfn2}. 
The calculation proceeded by evaluating the Bogoliubov coefficient in the near horizon limit outside 
the horizon. It was noted that crucial to obtaining the correct Hawking temperature is the choice of 
measure appropriate for eternal black holes. The measure was obtained from the DeWitt supermetric, 
$\gamma_{ab}$ on the configuration space $(\tau,R)$ and can be read directly from the Hamiltonian 
constraint
\beq
\gamma_{ab} = \left(\begin{matrix}
1 & 0 \cr
0 & \frac 1\cF\end{matrix}\right).
\eeq
It gives $\mu = 1/\sqrt{|\cF|}$, {\it i.e.,}
\beq
\langle \Psi_1,\Psi_2\rangle = \int \frac{dR}{\sqrt{|\cF|}} \Psi_1^\dagger \Psi_2
\eeq
and, via the hermiticity condition \eqref{meas}, the function 
\beq
A(R,F) = |\cF| \partial_R \ln (\sqrt{|\cF|})
\eeq
As long as $\cF\neq 0$ the Wheeler DeWitt equation can now be written as 
\beq
\left[\frac{\partial^2}{\partial \tau^2} \pm \frac{\partial^2}{\partial R_*^2} + B\right] \Psi = 0
\eeq
where the positive sign in the above equation refers to the exterior, while the negative 
sign refers to the interior and $R_*$ is defined by
\beq
R_* = \pm \int \frac{dR}{\sqrt{|\cF|}}.
\eeq

The second condition arises because we are describing a single shell in this simple quantum 
mechanical model of an eternal black hole and because $B(R,F)$ represents an interaction of 
the shell with itself. We simply demand there are no self interactions, {\it i.e.,} that 
$B(R,F)=0$. The quantum evolution is then described by the free wave equation in the interior, 
but by an elliptic equation in the exterior. This signature change has been noted in other 
models \cite{ki89,bk97} and occurs because of the behavior of $\cF$, which passes from positive 
outside the horizon to negative inside. For the black hole, it means that its wave function 
is supported in its interior. The spectrum will be determined by the proper radius, $L_h$, 
of the horizon,
\beq
L_h(M,\Lambda) = \int_0^{R_h} \frac{dR}{\sqrt{|\cF|}}
\label{propl}
\eeq
where $R_h$ is its area radius. If we extend the coordinate $R_*$ to range over $(-\infty,\infty)$, 
thereby avoiding any issues related to a boundary at the center \cite{va2}, this simple model of a 
quantum black hole effectively describes a dust shell in a ``box'' of radius $2L_h(M,l)$, which itself 
depends in a complicated way on its total ADM mass and the cosmological constant. The stationary states 
describe a spectrum of the form (reintroducing Planck's constant, $\hbar$)
\beq
\pA=\frac{n\Omega_n}{4\pi}~ (2G_d M) L_h = A_\text{Pl} \left(j+\frac 12\right)
\label{quant1}
\eeq
where $j$ is a whole number and $A_\text{Pl}=h G_d$ is the Planck area. Note that $\pA$ is not the 
horizon area although it has area dimension.  

It is not possible to give an analytical expression for $L_h$, but we will examine two 
interesting limits. First, defining $x=R/l$, we eliminate the mass, $M$, in 
favor of the horizon length, $x_h = R_h/l$, by using the fact that $x_h$ is a solution of $\cF=0$,
\beq
\frac{2G_d M}{l^{n-1}} = x_h^{n-1}\left[1+\frac{2x_h^2}{n(n+1)}\right],
\eeq
and express the proper length of the horizon as
\beq
L_h(x_h,l) = l \int_0^{x_h} \frac{dx}{\sqrt{\left(\frac{x_h}{x}\right)^{n-1}\left[1+\frac{2x_h^2}
{n(n+1)}\right]-\frac{2x^2}{n(n+1)}-1}}.
\eeq
Clearly, the dominant contribution to $L_h$ comes from the near horizon region, so when 
$2x_h^2 \ll n(n+1)$ (in the Schwarzschild limit) the integral may be approximated by
\beq
L_h(M) \approx l \int_0^{x_h} \frac{dx}{\sqrt{\left(\frac{x_h}{x}\right)^{n-1}-1}} = f(n) \frac{\pi R_h}
{n}.
\label{Lhlarge}
\eeq
where $R_h=(2G_dM)^{\frac 1{n-1}}$ is the area radius of the horizon and 
\beq
f(n) = \frac{n\Gamma\left(\frac 1{n-1}+\frac 12\right)}{\sqrt{\pi}\Gamma\left(\frac 1{n-1}\right)}
\eeq
is approximately one. Thus inserting $L_h$ into \eqref{quant1} shows that the spectrum can be described, 
in accordance with Bekenstein's conjecture, as a quantization of the horizon area ($A=\Omega_n R_h^n$) 
in half integer multiples of the Planck area according to 
\beq
\pA_j=f(n)\frac{A_j}4 = A_\text{Pl}\left(j+\frac 12\right),
\label{areaquant}
\eeq
up to some dimension dependent factors, all of which are contained in the function $f(n)$. We will 
show in the following section that the equispaced area spectrum predicted by our simplified model
of a quantum balck hole implies that the entropy, which is associated with (the natural logarithm of) 
the number of microstates compatible with a given macrostate of the black hole, obeys the 
Bekenstein-Hawking area law provided that the area quanta are assumed distinguishable. The entropy therefore 
also admits a discrete and evenly spaced spectrum.

On the other hand in the opposite limit, for which $2x_h^2 \gg n(n+1)$,
\beq
L_h(l) \approx l\int_0^{x_h} \frac{dx}{\sqrt{\frac{2}{n(n+1)}\left(\frac{x_h^{n+1}}{x^{n-1}}-x^2
\right)}} = \frac{l\pi} 2 \sqrt{\frac{2n}{(n+1)}}.
\eeq
Now inserting $L_h$ into \eqref{quant1} shows that this model predicts approximate {\it mass} quantization 
according to\footnote{Mass quantization and not area quantization occurs for the BTZ black 
hole for all $l$ \cite{va2}. There are no black holes in 2+1 dimensions without a cosmological 
constant.}
\beq
M_j = \frac{4h}{nl\Omega_n}\sqrt{\frac{n+1}{2n}} \left(j+\frac 12\right).
\label{massquant}
\eeq
We note that the gravitational constant, $G_d$, has completely disappeared from 
the quantization condition. The area spectrum is no longer equispaced and, assuming that the 
Bekenstein-Hawking area law still holds, neither is the entropy.\footnote{This statement is also 
true for the BTZ black hole.} We will argue that the 
Bekensitein-Hawking area law holds, provided that Bose statistics and not Boltzmann statistics 
are employed.

Both limits are verified explicitly when $n=3$, where an exact solution for $L_h$ is expressible in the 
simple form
\beq
L_h(x_h,l) = \sqrt{\frac 32}~ l \tan^{-1} \sqrt{\frac{2x_h^2}{3}\left(1+\frac{x_h^2}{3}\right)}.
\eeq
From this point of view the description of the black hole spectrum changes smoothly from area 
quantization in half integer units in the Schwarzschild limit, when $2x_h^2 \ll n(n+1)$, to mass 
quantization in half integer units when $2x_h^2 \gg n(n+1)$.

\section{Entropy}

Given the spectrum of $\pA$ we now ask for the origin of the degeneracy that leads to the black hole 
entropy. We examine here how the proposal in \cite{va2} may be extended to arbitrary dimensions. We 
treat the black hole as a single shell with the spectrum in \eqref{quant1}. However, this 
single shell is in fact the end state of many shells that have collapsed to form the black hole.
Regardless of their history, we assume that each of the shells then occupies only the levels of 
\eqref{quant1}, contributing some multiple of the Planck area to the total ``horizon area'', 
$\pA$, of the final state. A black hole microstate is thus a particular distribution of collapsed 
shells among the available levels. If the distribution of shells is such that $\cN_j$ shells occupy 
level $j$, the black hole's total ``horizon area'' becomes 
\beq
\pA = A_\text{Pl} \sum_j \left(j+\frac 12\right) \cN_j
\label{totA}
\eeq
and the (single shell) solution in \eqref{bh2} is to be interpreted as an excitation by $\cN=\sum_j \cN_j$ 
collapsed shells. The number of black hole microstates giving the ``area'' $\pA$ will depend on assumptions
concerning the degeneracy of the microstates. 

Since $\pA$ is approximately one quarter the horizon area when $2x_h^2 \ll n(n+1)$ according to 
\eqref{areaquant}, the sepctrum in \eqref{totA} represents the ``area ensemble''. Assuming 
the shells to be distinguishable, the number of states can be written in terms of the total number of ``area'' 
quanta, $\cQ$, and the total number of shells, $\cN$, as
\beq
\Omega(\cQ,\cN) = \frac{(\cN+\cQ-1)!}{(\cN-1)!\cQ!},
\eeq
where 
\beq
\cQ=\frac{\pA}{A_\text{Pl}} - \frac{\cN}2
\eeq
Holding $\pA$ fixed and extremizing the microcanonical entropy, $S_\text{micro}=k_B \ln \Omega$, 
with respect to the number of shells gives
\beq
S_\text{micro} = (2k_B f(n)\coth^{-1}\sqrt{5})\frac A{4 A_\text{Pl}},
\label{entmicro}
\eeq
which is in excellent (better than 91\% for any dimension) agreement with the Bekenstein-Hawking 
entropy. In addition to the exponential growth in the number of states, the area quantization in 
\eqref{totA} ensures that the entropy is effectively quantized in units of the Planck area, as 
originally proposed by Bekenstein \cite{bek73}. Loop Quantum Gravity (LQG) does not predict an 
evenly spaced area spectrum. However, very interesting recent work has unravelled a rich (and 
unexpected) band structure in the black hole degeneracy spectrum from LQG, leading to just such an 
{\it effective} equispacing of the entropy spectrum \cite{cor071,cor072,bar08}. 

The same result can be obtained in the canonical ensemble from the partition function
\beq
\tZ(\tbeta) = \sum_{\cN_1....\cN_j..} g(\cN_1,\ldots,\cN_j\ldots)\exp\left[-\tbeta A_\text{Pl} \sum_j 
\left(j+\frac 12\right)\cN_j\right]
\label{pf}
\eeq
in the saddle point approximation and subject to the constraint that $\sum_j \cN_j = \cN$. Note that here
$\tbeta$ is {\it not} the black hole temperature, being conjugate to $\pA$ and not the 
energy $M$ of the black hole. Distinguishability of the shells implies that $g(\cN_1,\ldots,\cN_j\ldots) =
\cN!/\cN_1!\ldots\cN_j!\ldots$ and gives
\beq
\tZ(\tbeta) = 2^{-\cN} \sinh^{-\cN}\left[\frac{\tbeta A_\text{Pl}}2\right]
\eeq
allowing us to compute the canonical entropy in the usual way as
\bea
S_\text{can} &=& k_B\left[\left.\ln \tZ(\tbeta)\right|_{\pA= -\partial\ln \tZ/\partial \tbeta} + \tbeta \pA
\right]\cr\cr 
&=& k_B\left[-\cN\ln 2 - \cN \ln \sinh\left[\coth^{-1}\left(\frac{2\pA}{\cN A_\text{Pl}}
\right)\right]+\frac{2\pA}{A_\text{Pl}}\coth^{-1} \left(\frac{2\pA}{\cN A_\text{Pl}}\right)\right].
\eea
Again extremizing with respect to the total number of shells yields
\beq
S_\text{can} = (2 k_B f(n)\coth^{-1}\sqrt{5})\frac{A}{4A_\text{Pl}},
\label{entSBH}
\eeq
in precise agreement with the microcanonical entropy in \eqref{entmicro}. It is a simple matter to 
show that had the shells been assumed indistinguishable then the entropy would depend on the 
square-root of the area \cite{Kolland}. The fact that the area degrees of freedom must be treated as
distinguishable runs contrary to our intuition for elementary particles in quantum field theory and 
calls into question whether ``area'' is a fundamental quantity in quantum gravity. 

In the opposite limit, $2x_h^2 \gg n(n+1)$, it is the mass and not the area that 
is quantized in half integer units according to \eqref{massquant}. We therefore consider the 
``energy ensemble'', in which the total mass of the black hole may be given as 
\beq
M = \frac{4h}{nl\Omega_n} \sqrt{\frac{n+1}{2n}}\sum_j \left(j+\frac 12\right) \cN_j,
\eeq
The partition function is (now $\beta$ is the inverse temperature of the black hole)
\beq
Z(\beta) = \sum_{\{\cN_1,\ldots,\cN_j,\ldots\}}g(\cN_1,\ldots,\cN_j,\ldots) \exp
\left[-\frac{4h\beta}{nl\Omega_n} \sqrt{\frac{n+1}{2n}}\sum_j\left(j+\frac 12\right) \cN_j\right]
\eeq
and this time the area dependence of the Bekenstein-Hawking entropy cannot be obtained by treating the 
shells as distinguishable. Instead we show below that to recover the correct area dependence they
must be treated as indistinguishable. Taking $g(\cN_1,\ldots,\cN_j,\ldots)=1$, we must then evaluate
\beq
Z(\xi) = \prod_{j=0}^\infty\left[1-e^{-\xi(2j+1)}\right]^{-1}
\eeq
with 
\beq
\xi=\frac{2h \beta}{nl\Omega_n}\sqrt{\frac{n+1}{2n}}.
\label{xivalue}
\eeq
Exploiting the well known high/low temperature duality of the Bose partition function \cite{hr18}, 
we obtained in \cite{va2} a relationship which explicitly relates the partition function at small 
argument to the partition function at large argument,
\beq
Z(\xi) = \prod_{j=0}^\infty \left[1-e^{-\xi(2j+1)}\right] = \frac 1{\sqrt{2}}e^{\frac{\pi^2}{12\xi}
+\frac{\xi}{24}}[Z(2\pi^2/\xi)]^{-1}.
\label{duality}
\eeq
We will be interested in determining the partition function when $\xi\ll 1$. The value of 
$Z(2\pi^2/\xi)$ depends on assumptions concerning ``ground state'' of the system. If we think of  
$Z(2\pi^2/\xi)$ as the partition function in the Schwarzschild limit, for which we derived 
the entropy in \eqref{entSBH}, we find 
\beq
Z\left(\frac{2\pi^2}\xi\right) \approx \exp\left[-\frac{f(n)\coth^{-1}\sqrt{5}\Omega_n}{2(n-1)h G}
\left(\frac{\Omega_n(n-1)\pi^2 l}{f(n)\coth^{-1}\sqrt{5}}\sqrt{\frac{2n}{n+1}}\right)^n\xi^{-n}\right] 
\stackrel{\text{def}}{=} e^{-\sigma_n \xi^{-n}}
\eeq
and from \eqref{duality} it follows that 
\beq
\ln Z(\xi) = -\frac 12 \ln 2 + \frac{\pi^2}{12\xi}+\frac{\xi}{24}+\frac{\sigma_n}{\xi^n},
\eeq
For small values  of $\xi$ and $n\geq 2$ the right hand side is dominated by the last term,
therefore
\beq
M = - \frac{\partial \ln Z}{\partial \beta} \approx n \sigma_n \left(\frac{2h}
{nl\Omega_n}\sqrt{\frac{n+1}{2n}}\right)^{-n} \beta^{-n-1}
\eeq
and we obtain the canonical entropy in the saddle point approximation as
\beq
S_\text{can}= \left(\ln Z + \beta M\right)_{M=-\partial \ln Z/\partial\beta} = 
(8\pi^2)^{\frac n{n+1}}\left(\frac{(n-1)}{2(n+1)f(n)\coth^{-1}\sqrt{5}}\right)^{\frac{n-1}{n+1}} 
\frac{A}{4A_\text{Pl}}
\eeq
where, in the limit of large black holes, 
\beq
A = \Omega_n \left[n(n+1) l^2 (GM)\right]^\frac{n}{n+1}
\eeq
is just the area of the horizon. The fact that we have recovered an area law is critical for 
consistency because we are quantizing Einstein's gravity, whose equivalence with the Bekenstein-Hawking
area law was established in \cite{jacobson95}. It justifies our use of the Schwarzschild limit as
the ``ground state'' of the system. We may now examine $\xi$ more carefully by expressing the inverse 
temperature $\beta$ in terms of the horizon surface gravity, 
\beq
\frac \beta l = \frac{2\pi}{l\kappa} = 4\pi\left[\frac{n-1}{x_h} + \frac{2x_h}{n}\right]^{-1}.
\eeq
Viewed as a function of $x_h$, the right hand side has an absolute maximum value at 
$2x_{h,\text{min}}^2=n(n-1)$. 
This implies that $\xi$ is bounded from above and taking $\xi\ll 1$ is equivalent to the condition
$l\kappa \gg 1$, which clearly holds in the limit we are considering.

\section{Discussion}

In this paper we have examined two limits of a single model of quantum AdS black holes.  There 
are factor ordering ambiguities in our description, but ambiguities are present in all
approaches to date. Choosing factor orderings different from the one we have used would lead 
to different spectra and so alternate descriptions of the black hole states. We do not know how 
or even if the descriptions would be connected.  The factor ordering employed in this paper is 
consistent with our description of Hawking radiation in \cite{va4}. It does not arise for genuine 
collapse scenarios, where the matter distribution is taken to be smooth over some interval of the 
ADM label coordinate, $r$. There, the diffeomorphism constraint uniquely fixes the factor ordering 
as well as the measure on the Hilbert space \cite{kmhv05}. It is therefore possible that, once 
the quantum collapse process leading to the final state is more fully understood, this uniqueness 
will carry over to a unique description of the end state. This seems to be a worthwhile direction 
for future research.

We have shown that the canonical theory (with the chosen factor ordering) predicts a linear excitation 
spectrum of a certain quantity, {\it i.e.,} $\pA(M,l)$. However, $\pA(M,l)$ has very different 
interpretations in the two limits that are also addressed by other approaches to quantum gravity. 
As the cosmological constant vanishes, $\pA(M,l)$ turns into the horizon area and our 
model yields equispaced area levels. This spectrum agrees with the predictions of LQG, modulo a further 
degeneracy due to a quantum number that is responsible for the intrinsic geometry 
\cite{cor071,cor072,bar08}, provided the horizon 
area is large. In the limit of a large cosmological constant, $\pA(M,l)$ is the black hole mass-energy. 
This spectrum coincides with the string and AdS/CFT approaches.  Although the area law 
was shown long ago to be equivalent to Einstein's gravity \cite{jacobson95} both with and without a 
cosmological constant, to extract the area dependence of the black hole entropy from the quantization 
of Einstein's theory it is necessary to use different statistics in each limit: Boltzmann in the first 
limit and Bose in the second. 

In the Schwarzschild limit, the entropy spectrum of the black hole is equispaced in our model. While 
this is not na\"ively apparent in LQG, recent numerical work has shown the existence of a band 
structure in the degeneracy that leads to an effective equispaced quantization of the entropy 
\cite{cor071,cor072,bar08}. It would be interesting, but is not clear at this time, if simplified 
models such as the one we have worked with in this paper can shed some light on this issue. 

One feature of our solutions in the opposite limit is the independence of the spectrum on the 
gravitational constant, $G_d$, in all dimensions. According to \eqref{xivalue} and \eqref{duality}, 
the gravitational constant enters in the thermodynamic description via a duality which connects 
the partition functions 
in the two limits. In more than three dimensions, these limits are separated by a first order phase 
transition that occurs when $2x_h^2 = n(n-1)$ \cite{hawpag83,das02} so they are two different phases 
of the same thermodynamic system, the former with negative heat capacity and the latter with positive 
heat capacity.\footnote{In 2+1 dimensions, the picture of the black hole as a stationary state built 
out of collapsed shells remains but only mass quantization occurs and one recovers the area law using 
Bose statistics for all values of the cosmological constant \cite{va2}. The heat capacity is 
positive throughout and no phase transition occurs.} The limit $2x_h^2\gg 2n(n+1)$ is more in keeping 
with quantum field theory both because the heat capacity is positive and because the fundamental 
``particles'' must be assumed indistinguishable. The phase transition may describe a change in the 
nature of the fundamental degrees of freedom from ``field theoretic'' to ``geometric''. 

The states we have obtained are valid for all values of $x_h$, yet they are eigenstates of different 
physical quantities in each limit and must be treated as distinguishable or indistinguishable depending 
on the physical interpretation of the spectrum. This calls for a better understanding. A microscopic 
description of the thermodynamics in between these two limits should be illuminating.
%%\vfill\eject
\bigskip\bigskip

\noindent{\bf Acknowledgements}
\bigskip

\noindent LCRW was supported in part by the U.S. Department of Energy Grant No. DE-FG02-84ER40153. 

\end{document}